\documentclass[12pt,a4paper,fleqn]{article}
\usepackage{bbm}
\usepackage{amsfonts}
\usepackage[fleqn]{amsmath}
\usepackage{amsmath}
\usepackage{mathrsfs}
\usepackage{amssymb}
\usepackage{latexsym}
\usepackage{stmaryrd}
\usepackage{makeidx}
\usepackage[all]{xy}
\usepackage{CJK}
\usepackage{indentfirst}
\usepackage{tikz}
\usepackage{array}
\usepackage{graphicx}
\usepackage{epsfig}
\usepackage{subfigure}
\usepackage{color}
\usepackage{cite}
\usepackage[percent]{overpic}
\usepackage[top=1in, bottom=1in, left=0.5in, right=0.5in]{geometry}
\usepackage[colorlinks, linkcolor=blue, anchorcolor=blue,citecolor=blue]{hyperref}
\date{}

\title{\Large \textbf{Obtaining nonvanishing $\theta_{13}$ with constrained neutrino Yukawa matrix and implications for flavor model buildings}}
\author{Ya Zhao  $^{1, }$ \footnote{\emph{E-mail address}: zhaoya@mail.ustc.edu.cn}
\bigskip
\\
{ $^{1}$ {\footnotesize
 \it Department of Modern Physics, University of Science and Technology  of China, Hefei, Anhui 230026, China.}}}
\begin{document}
\normalsize
\maketitle
\begin{abstract}
  Assuming a diagonal Majorana neutrino mass matrix, we investigate the neutrino Yukawa textures which lead to a non-zero reactor mixing angle
  $\theta_{13}$. The neutrino effective coupling matrix $\kappa^{eff}$ is pre-diagonalized by a constant mixing pattern $V_{\nu}$ with a
  vanishing $\theta^{\nu}_{13}$. The resulting pre-diagonal symmetrical matrix $\kappa$ is set to be four texture zeros with two types of
  off-diagonal elements nonzero, which is $\kappa_{13}$ and $\kappa_{23}$, respectively. With the expectation of simple textures we thoroughly
  classify the linear combinations, $\alpha_{i}$, $\beta_{i}$ and $\gamma_{i}$ of Yukawa elements $\lambda_{ij}$ in a same row, according
  to the values vanishing or not. Each set of the classifications can lead to a Yukawa texture which may have implications for the discrete
  flavor model buildings. We also present a model based on $A_{4}$ according to one set of the constraints on the three combinations with
  a specific choice of a coefficient in Yukawa texture.
\end{abstract}
\section{Introduction}\label{S1}
Despite of being well compatible with experimental measured large solar and atmospherical neutrino mixing angles,
the neutrino mixing patterns $V_{\nu}$ with vanishing reactor neutrino mixing angle $\theta^{\nu}_{13}$, such as the \emph{ansatz}
Tri-Bimaximal~\cite{HPS/2002}, and Golden-ratio~\cite{GR/2007,Rode/2009a,Adul/2009d} etc., are now ruled out by
rigidly experimental results~\cite{dyb/2012,RENO/2012}. For obtaining a sizable lepton mixing angle $\theta^{\ell}_{13}$,
it is necessary to induce corrections to the patterns, especially those from sizable mixing in charged lepton are most motivated in the GUT context.
The scenario can be realized in the models based on discrete non-Abelian family symmetries, especially those combined with GUT gauge symmetries.
In the models of this kind the leptonic mixing PMNS matrix is given by $V_{\ell}=V^{\dag}_{e}V_{\nu}$, in which a typical $V_{e}$ in charged lepton
sector has a sizable $\theta^{e}_{12}\simeq\lambda_{c}$, with $\lambda_{c}\approx0.225$ being the Cabibbo angle. The mixing matrix $V_{\nu}$
in neutrino sector owns large $\theta^{\nu}_{12}$ and $\theta^{\nu}_{23}$. Then $V_{\ell}$ yields the empirical relation
$\theta^{\ell}_{13}\simeq\lambda_{c}/\sqrt{2}$ when $\theta^{\nu}_{23}=\pi/4$, which coincides with the experiment results.

On the other side, in order to reduce the number of parameters in effective light neutrino mass matrix $m_{\nu}$ and reveal the correlation 
between mixing angles and masses, the appealing schemes of texture
zeros~\cite{Frampton/2002tz,Xing/2002tz,Randhawa/2006tz,Merle/2006tz,DevS/2007tz,Kumar/2011tz,Ludl/2012tz,Grimus/2013tz,Blankenburg/2013tz,
Fritzsch/2011tz,Hirsch/2007A4,ZhouS/2016Tz,DevS/2012Tz,Deepthi/2012Tz,LiaoJ/2013Tz,Ludl/2014sTz,Berger/2001Tz}, 
cofactor zeros~\cite{Cozero/2013}, vanishing minors~\cite{Lavoura/2005vm,Lashin/2008vm,DevS/2011vm,ArakiT/2012vm,Verma/2012vm,
DevS/2010vm,LiaoJ/2014vm} and hybrid textures~\cite{KanekoS/2005Hyb,DevS/2010Hyb,Goswami/2010Hyb} etc., are well studied.
Most of them focus on texture zeros in matrix $m_{\nu}$ itself or Dirac mass matrix as well as Majorana mass matrix.
These textures can be realized as well within the context of proper Abelian flavor symmetries. Nonetheless the realization of the textures
in non-Abelian discrete flavor symmetries is not an easy task for model buildings, only a few of such models have been appeared in
literatures~\cite{Hirsch/2007A4,ZhouS/2016Tz}. The spirit of texture zeros, however, inspire us to seek another strategy to build non-Abelian
flavor models. Compared with the original mass matrix $m_{\nu}$, the matrix $m'_{\nu}=V^{T}_{\nu}m_{\nu}V_{\nu}$ (refer as to pre-diagonal matrix) 
with texture zeros is an alternative to investigate the correlative relations between mixing parameters and masses. Besides the potential reduction
of parameters may reveal new relations between mass entries in either Dirac type or Majorana type. The minimal scenario to
obtain a non-zero $\theta^{\ell}_{13}$ is that only (13) or (23) elements (as well as (31) or (32) elements) are nonzero
in the pre-diagonal mass matrix. Hence $m'_{\nu}$ with four texture zeros can be thoroughly diagonalized by an extra small rotation
$\delta V$ in (13) or (23) complex plane. The leptonic mixing matrix $V_{\ell}=V_{\nu}\delta V$ will lead to the desired 
nonvanishing $\theta^{\ell}_{13}$.

In this paper $V_{\nu}$ is taken as tri-bimaximal mixing, we investigate the texture of neutrino Yukawa matrix $\lambda$ in flavor basis 
where both the charged leptons mass matrix $m_{e}$ and heavy majorana neutrinos $\mu$ are diagonal. The effective neutrino coupling 
matrix $\kappa^{eff}=\lambda^{T}\mu^{-1}\lambda$ (which is equal to $m_{\nu}/v^2$ with $v$ the Higgs vev) is pre-diagonalized by $V_{\nu}$, which 
gives the pre-diagonal coupling $\kappa$. By taking $\kappa_{13}$ and $\kappa_{23}$ vanishing respectively, the constraint conditions are deduced 
from the linear combinations, $\alpha_{i}$, $\beta_{i}$ and $\gamma_{i}$, of the Yukawa elements $\lambda_{ij}$ in a same row. 
According to the constraints we obtain a simple Yukawa matrix as example, which can be realized in $A_{4}$ family symmetry model.

The paper is organized as follows. In Sec.~\ref{S2} we take $V_{\nu}$ to be tri-bimaximal mixing and discuss the two minimal scenarios of 
pre-diagonal coupling matrix $\kappa$, which lead to a nonvanishing $\theta^{\ell}_{13}$. In Sec.~\ref{S3} we 
classify the constraint conditions of $\alpha_{i}$, $\beta_{i}$ and $\gamma_{i}$ for obtaining a nonvanishing $\kappa_{13}$ and $\kappa_{23}$ 
in detail. We also discuss the implications for the flavor model buildings according to the constraints, and a model based on $A_{4}$ family symmetry
is presented as an example. Finally Sec.~\ref{S4} is the conclusion.

\section{Basic aspects}\label{S2}
In the Lagrangian of standard model (SM) there is no neutrino mass term due to the absent of right-handed neutrinos. For generating nonzero neutrino
masses in a minimal extended SM one can add right-handed fields $\nu_{R}$ like the other right-handed fields which belong to $SU(2)_{L}$ singlets.
Without loss of generality we consider three heavy Majorana right-handed neutrinos $\nu_{R}$ are introduced in the minimal extended SM, 
then the relevant Lagrangian in lepton sector reads
\begin{equation}\label{eq:Lagrangian}
  -\mathcal{L}=Y_{e}\bar{\ell}_{L}\phi\ell_{R}+\lambda\bar{\ell}_{L}\tilde{\phi}\nu_{R}+\frac{1}{2}\mu\overline{\nu^{C}_{R}}\nu_{R}+h.c.,
\end{equation}
where $\phi$ is the $SU(2)_{L}$ Higgs boson doublet with $\tilde{\phi}\equiv i\tau_{2}\phi^{\ast}$ and the family indexes are omitted. Here 
$\nu^{c}=C\bar{\nu}^{T}$ and $C$ is the charge conjugation matrix, and $\ell_{L}$ and $\ell_{R}$ are the ordinary lepton left handed doublet 
and right handed singlet of $SU(2)_{L}$, respectively. We work in the flavor basis that both the charged lepton mass matrix and Majorana 
mass matrix are diagonal, i.e., $m_{e}=Y_{e}v=\textrm{Diag}\{y_{e},y_{\mu},y_{\tau}\}v$ with $v=\langle\phi\rangle=246\textrm{GeV}$ being 
the vev of the neutral component of Higgs boson doublet after the electroweak symmetry breaking, 
and $\mu=\textrm{Diag}\{\mu_{1},\mu_{2},\mu_{3}\}$. The neutrino Yukawa matrix is in general a complex matrix which would encode the leptonic 
flavor structure as following
\begin{equation}\label{eq:Yukawa}
  \lambda=\left(
            \begin{array}{ccc}
              \lambda_{11} & \lambda_{12} & \lambda_{13} \\
              \lambda_{21} & \lambda_{22} & \lambda_{23} \\
              \lambda_{31} & \lambda_{32} & \lambda_{33} \\
            \end{array}
          \right),
\end{equation}
thus the Dirac neutrino mass matrix $m_{D}=\lambda v$. Accordingly the Lagrangian for lepton masses is
\begin{equation}\label{eq:Lagrangianmass}
  -\mathcal{L}_{mass}=(\overline{e_{L}},\overline{\mu_{L}},\overline{\tau_{L}})m_{e}(e_{R},\mu_{R},\tau_{R})^{T}
  +(\overline{\nu_{eL}},\overline{\nu_{\mu L}},\overline{\nu_{\tau L}})m_{D}(\nu_{1R},\nu_{2R},\nu_{3R})^{T}
  +\frac{1}{2}\mu_{i}\overline{\nu^{C}_{iR}}\nu_{iR}+h.c.,
\end{equation}
then the light effective left-handed Majorana neutrino mass term is given by seesaw mechanism after integrating out the heavy right-handed neutrinos
\begin{equation}\label{eq:Lagrangianlight}
  -\mathcal{L}_{mass}=(\overline{e_{L}},\overline{\mu_{L}},\overline{\tau_{L}})m_{e}(e_{R},\mu_{R},\tau_{R})^{T}+
  \frac{1}{2}(\overline{\nu_{eL}},\overline{\nu_{\mu L}},\overline{\nu_{\tau L}})m_{\nu}(\nu^{c}_{eL},\nu^{c}_{\mu L},\nu^{c}_{\tau L})^{T}+h.c..
\end{equation}
The light neutrino mass matrix $m_{\nu}$ in the basis is $m_{\nu}=-k^{eff}v^{2}$ where the effective neutrino coupling matrix $\kappa^{eff}$
is produced by type-I seesaw
\begin{equation}\label{eq:kappa}
  \kappa^{eff}=\lambda^{T}\mu^{-1}\lambda.
\end{equation}
The eigenvalues of $\kappa^{eff}$ is obtained by the unitary transformation $V_{\ell}$: $\kappa^{diag}=V^{T}_{\ell}\kappa^{eff}V_{\ell}$. Usually
$V_{\ell}=V^{\dag}_{e}V_{\nu}$ is the very lepton mixing PMNS matrix. In the flavor basis $V_{e}$ is a unit matrix $V_{e}=\textbf{1}$, and $V_{\nu}$ is
often taken as a pattern with large $\theta^{\nu}_{12}$ and $\theta^{\nu}_{23}$ and vanishing $\theta^{\nu}_{13}$. Under the condition we may obtain
the following symmetrical coupling matrix through the unitary transformation
\begin{equation}\label{eq:kappaprediagonal}
  \kappa=V^{T}_{\nu}\kappa^{eff} V_{\nu}=\left(
                             \begin{array}{ccc}
                               \kappa_{11} & \kappa_{12} & \kappa_{13} \\
                               \kappa_{12} & \kappa_{22} & \kappa_{23} \\
                               \kappa_{13} & \kappa_{23} & \kappa_{33} \\
                             \end{array}
                           \right).
\end{equation}
In the following $\kappa$ is referred to as the pre-diagonal coupling matrix, and when we say one off-diagonal element $\kappa_{ij}$ vanishing or not
means the same constraints on its symmetrical one $\kappa_{ji}$ since they are equal to each other. We note that in general the neutrino mixing
matrix $V_{\nu}$ is not identical to the PMNS lepton mixing matrix. Taking tribimaximal mixing $V_{\nu}=V_{HPS}$~\cite{HPS/2002} as example,
the pattern reads
\begin{equation}\label{eq:VHPS}
  V_{\nu}=\left(
            \begin{array}{ccc}
              \sqrt{\frac{2}{3}} & \frac{1}{\sqrt{3}} & 0 \\
              -\frac{1}{\sqrt{6}} & \frac{1}{\sqrt{3}} & -\frac{1}{\sqrt{2}} \\
              -\frac{1}{\sqrt{6}} & \frac{1}{\sqrt{3}} & \frac{1}{\sqrt{2}} \\
            \end{array}
          \right),
\end{equation}
then we can easily get each elements $\kappa_{ij}$ in terms of Yukawa elements $\lambda_{ij}$ and Majorana masses $\mu_{i}$. For the diagonal
elements of $\kappa$ we have
\begin{equation}\label{eq:kappaii}
  \kappa_{11}=\frac{1}{6}\sum^{3}_{i=1}\frac{\alpha^{2}_{i}}{\mu_{i}},\qquad
  \kappa_{22}=\frac{1}{3}\sum^{3}_{i=1}\frac{\beta^{2}_{i}}{\mu_{i}},\qquad
  \kappa_{33}=\frac{1}{2}\sum^{3}_{i=1}\frac{\gamma^{2}_{i}}{\mu_{i}},
\end{equation}
and the off-diagonal elements are
\begin{equation}\label{eq:kappaij}
  \kappa_{12}=\frac{1}{3\sqrt{2}}\sum^{3}_{i=1}\frac{\alpha_{i}\beta_{i}}{\mu_{i}},\qquad
  \kappa_{13}=-\frac{1}{2\sqrt{3}}\sum^{3}_{i=1}\frac{\alpha_{i}\gamma_{i}}{\mu_{i}},\qquad
  \kappa_{23}=-\frac{1}{2\sqrt{6}}\sum^{3}_{i=1}\frac{\beta_{i}\gamma_{i}}{\mu_{i}},
\end{equation}
in which the parameters $\alpha_{i}$, $\beta_{i}$ and $\gamma_{i}$ are linear combinations of elements $\lambda_{ij}$ ($j=1,2,3$) in a same
row of $\lambda$ as follows
\begin{equation}\label{eq:alpha}
  \alpha_{i}=2\lambda_{i1}-\lambda_{i2}-\lambda_{i3},\qquad
  \beta_{i}=\lambda_{i1}+\lambda_{i2}+\lambda_{i3},\qquad
  \gamma_{i}=\lambda_{i2}-\lambda_{i3}.
\end{equation}
For the case that $\theta^{\nu}_{12}$ is taken a different value and $\theta^{\nu}_{23}$ remains maximum in $V_{\nu}$, the analysis is analogously.
The only change is that the coefficients ahead of $\lambda_{i1}$ in $\alpha_{i}$ and $\beta_{i}$ are replaced by $\sqrt{2}\cot\theta^{\nu}_{12}$
and $\sqrt{2}\tan\theta^{\nu}_{12}$, respectively. The corresponding constant coefficients of $\kappa_{ij}$ in eq.~\eqref{eq:kappaii}
and eq.~\eqref{eq:kappaij} should be altered as well. In addition, in the minimal seesaw where the heavy Majorana neutrino mass matrix is
$\mu=\textrm{Diag}\{\mu_{1},\mu_{2}\}$ and Yukawa matrix $\lambda$ is a $2\times3$ matrix, all the elements $\kappa_{ij}$
in eq.~\eqref{eq:kappaii} and~\eqref{eq:kappaij} in corresponding pre-diagonal coupling matrix $\kappa$ are only summed from 1 to 2.

With all the off-diagonal elements vanishing one may in principle constrain the Yukawa textures which lead to the famous tribimaximal mixing
$V_{\nu}=V_{HPS}$. In order that the sizable lepton mixing angle $\theta^{\ell}_{13}$ can be produced in neutrino sector, there are two minimal
schemes in which only one off-diagonal element is non-vanishing in the pre-diagonal coupling matrix. For clarity we write the two minimal scenarios
with four texture zeros as following
\begin{equation}\label{eq:kappai3}
  \kappa=\left(\begin{array}{ccc}
     \kappa_{11} & 0 & \kappa_{13} \\
     0 & \kappa_{22} & 0 \\
     \kappa_{13} & 0 & \kappa_{33} \\
     \end{array}
     \right), \quad\textrm{or}\quad
  \left(
  \begin{array}{ccc}
  \kappa_{11} & 0 & 0 \\
  0 & \kappa_{22} & \kappa_{23} \\
  0 & \kappa_{23} & \kappa_{33} \\
  \end{array}
  \right)
\end{equation}
Taking the first case that $\kappa_{13}$ is nonvanishing as example, one can easily diagonalize $\kappa$ in eq.~\eqref{eq:kappai3} with an extra
unitary transformation $\delta V$ of the form
\begin{equation}\label{eq:deltaV}
  \delta V_{13}=\left(
             \begin{array}{ccc}
               \cos\frac{\vartheta}{2} & 0 & \sin\frac{\vartheta}{2}e^{-i\delta} \\
               0 & 1 & 0 \\
               -\sin\frac{\vartheta}{2}e^{i\delta} & 0 & \cos\frac{\vartheta}{2} \\
             \end{array}
           \right),
\end{equation}
The eigenvalues of $\kappa$ are solved as follows
\begin{eqnarray}\label{eq:kappavalue}
  & & \kappa_{1} = \kappa_{11}\cos^{2}\frac{\vartheta}{2}+\kappa_{33}\sin^{2}\frac{\vartheta}{2}e^{2i\delta}
  -\kappa_{13}\sin\vartheta e^{i\delta}, \nonumber\\
  & & \kappa_{2} = \kappa_{22}, \nonumber\\
  & & \kappa_{3} = \kappa_{11} \sin^{2}\frac{\vartheta}{2}e^{-2i\delta}+\kappa_{33}\cos^{2}\frac{\vartheta}{2}
  +\kappa_{13}\sin\vartheta e^{i\delta},
\end{eqnarray}
where $\vartheta$ is given by
\begin{equation}\label{eq:vartheta}
  \tan\vartheta=\frac{2\kappa_{13}e^{i\delta}}{\kappa_{33}e^{2i\delta}-\kappa_{11}},
\end{equation}
and the Dirac CP violation phase $\delta$ has to be the following form
\begin{equation}\label{eq:delta}
  \delta=\frac{i}{2}\ln\frac{\kappa^{\ast}_{13}\kappa_{33}+\kappa_{13}\kappa^{\ast}_{11}}{\kappa_{13}\kappa^{\ast}_{33}+\kappa^{\ast}_{13}\kappa_{11}},
\end{equation}
to keep $\tan\vartheta$ in eq.~\eqref{eq:vartheta} real. In the similar way we can obtain the corresponding results for the second case.
The PMNS matrix is then defined as
\begin{equation}\label{eq:VPMNS}
  V_{\ell}=V_{\nu}\delta VP_{\nu}, \quad P_{\nu}=\textrm{diag}\{e^{i\rho},e^{i\sigma},1\},
\end{equation}
which yields the final lepton mixing angles $\theta^{\ell}_{ij}$ in terms of $\vartheta$ and $\delta$ as follows
\begin{eqnarray}
  \sin^{2}\theta^{\ell}_{13} &=& \frac{2}{3}\sin^{2}\frac{\vartheta}{2}, \nonumber\\
  \sin^{2}\theta^{\ell}_{12} &=& \frac{1}{3-2\sin^{2}\frac{\vartheta}{2}}, \nonumber\\
  \sin^{2}\theta^{\ell}_{23} &=& \frac{2+\cos\vartheta+\sqrt{3}\sin\vartheta\cos\delta}{4+2\cos\vartheta},
\end{eqnarray}
and $\rho$ and $\sigma$ in $P_{\nu}$ are Majorana CP violating phases. Note that in order that the mixing angle $\theta^{\ell}_{13}$
(as well as $\theta^{\ell}_{12}$ and $\theta^{\ell}_{23}$) can be compatible with experimental bounds, the potential fine tuning on the
order of magnitude in Yukawa elements or Majorana masses are allowed by hand or other possible allowed dynamical mechanism. The question is
beyond the scope of present discussions. The direct constraint on $\vartheta$ can be, however, estimated by the global fit
data~\cite{Garcia/2015fit}. For simplicity we only take $\theta^{\ell}_{13}$ as a demonstration, the best fit at $1\sigma$ level gives that
\begin{equation}\label{eq:theta13fit}
  \sin^{2}\theta^{\ell}_{13}=0.0218\pm0.0010(\textrm{NO}),\qquad 0.0219^{+ 0.0011}_{-0.0010}(\textrm{IO}),
\end{equation}
where NO (IO) represents normal (inverted) order of the light neutrino masses. Hence $\vartheta$ is given by
\begin{equation}\label{eq:varthetanum}
  \sin^{2}\frac{\vartheta}{2}=0.0327\pm0.0015(\textrm{NO}),\qquad 0.03285^{+0.00165}_{-0.0015}(\textrm{IO}),
\end{equation}
accordingly one can restrict the elements $\kappa_{ij}$ in eq.~\eqref{eq:vartheta}.

\section{Realization of the texture of $\kappa$}\label{S3}
In order to generate nonvanishing $\kappa_{13}$ or $\kappa_{23}$ in eq.~\eqref{eq:kappaij}, we have to check the correlation between Yukawa
elements $\lambda_{ij}$. For sake of simplicity we first enforce all the phases $\phi_{ij}=\arg(\lambda_{ij})$, of elements $\lambda_{ij}$
with the same row index $i$, to be the same. At the same time we also assume $\kappa_{ij}$ to be vanishing if and only if the
three terms therein are all vanishing. In the case we may focus on the simplest constraints on the Yukawa elements, i.e., constraints from
$\alpha_{i}$, $\beta_{i}$ and $\gamma_{i}$, which can be realized in some specific discrete flavor models. Generally there is at least one
nonvanishing $\alpha_{i}$ ($i=1,2,3$) to guarantee the diagonal element $\kappa_{11}$ nonvanishing. The same requests for $\beta_{i}$ and $\gamma_{i}$
should be satisfied as well. Next we would classify the conditions for obtaining the $\kappa_{13}$ or $\kappa_{23}$ non-vanishing based on the
above assumptions.
\subsection{Classification}
Next we discuss the case that all diagonal elements $\kappa_{ii}$ are non-vanishing which guarantee the mass eigenvalues are nonzero ones. There are
two kinds of classifications of $\kappa_{11}$ in terms of heavy Majorana masses with one or two nonvanishing $\alpha_{i}$. The third kind with three
nonzero $\alpha_{i}$, however, will lead to all $\beta_{i}$ vanishes for keeping $\kappa_{12}=0$. Then the diagonal element $\kappa_{22}$ is zero
as well which is incompatible with the requirement of $\kappa_{ii}$ always nonzero. We drop the case in the following analysis.

The detailed constrains from the request of $\kappa_{ij}$ to be vanishing or non-vanishing are presented in~\hyperref[ta:tab1]{Table 1}
($\kappa_{13}\neq0$) and~\hyperref[ta:tab2]{Table 2} ($\kappa_{23}\neq0$). Note that neither the order of magnitudes of Yukawa elements
$\lambda_{ij}$ nor those of Majorana masses $\mu_{i}$ are fixed in the analysis. The restrictive conditions
$\gamma_{i}\neq0/\gamma_{j}\neq0/\gamma_{i,j}\neq0$ ($i\neq j$) in~\hyperref[ta:tab2]{Table 2} denote that
$(\gamma_{i}\neq0,\gamma_{j}=0)/(\gamma_{i}=0,\gamma_{j}\neq0)/\gamma_{i,j}\neq0$.
\begin{table}[!htbp]\footnotesize
  \centering
  \caption{Classifications of the constraint conditions on $\kappa_{13}\neq0$ and $\kappa_{12}=\kappa_{23}=0$.}\label{ta:tab1}
  \resizebox{\textwidth}{!}{
  \begin{tabular}{|c|c|c|c|c|c|c|c|}
  \hline
  \hline
  Entries & $\kappa_{11}\neq0$ & $\kappa_{12}=0$ & $\kappa_{13}\neq0$ & $\kappa_{22}\neq0$ & $\kappa_{23}=0$ & $\kappa_{33}\neq0$ & case \\
  \hline
   & & & & $\beta_{2}\neq0,\beta_{3}=0$ & $\gamma_{2}=0$ & $\gamma_{3}\neq0/=0$ & a \\
  I& $\alpha_{1}\neq0,\alpha_{2,3}=0$ &  $\beta_{1}=0$& $\gamma_{1}\neq0$ & $\beta_{2}=0,\beta_{3}\neq0$
  & $\gamma_{3}=0$ & $\gamma_{2}\neq0/=0$ & b \\
   & & & & $\beta_{2,3}\neq0$ & $\gamma_{2,3}=0$ & $\gamma_{1}\neq0$ & c \\
  \hline
     & & & & $\beta_{1}\neq0,\beta_{3}=0$ & $\gamma_{1}=0$ & $\gamma_{3}\neq0/=0$ & a \\
  II & $\alpha_{2}\neq0,\alpha_{1,3}=0$ & $\beta_{2}=0$ &  $\gamma_{2}\neq0$ & $\beta_{1}=0,\beta_{3}\neq0$
  & $\gamma_{3}=0$ & $\gamma_{1}\neq0/=0$ & b \\
     & & & & $\beta_{1,3}\neq0$ & $\gamma_{1,3}=0$ & $\gamma_{2}\neq0$ & c \\
  \hline
     & & & & $\beta_{1}\neq0,\beta_{2}=0$ & $\gamma_{1}=0$ & $\gamma_{2}/\neq0=0$ & a\\
  III & $\alpha_{3}\neq0,\alpha_{1,2}=0$ & $\beta_{3}=0$ & $\gamma_{3}\neq0$ & $\beta_{1}=0,\beta_{2}\neq0$
  & $\gamma_{2}=0$ & $\gamma_{1}\neq0/=0$ & b \\
     & & & & $\beta_{1,2}\neq0$ & $\gamma_{1,2}=0$ & $\gamma_{3}\neq0$ & c\\
  \hline
  \hline
     & & & $\gamma_{1}\neq0,\gamma_{2}=0$ & & & $\gamma_{1}\neq0$ & a \\
  IV & $\alpha_{1,2}\neq0,\alpha_{3}=0$ & $\beta_{1,2}=0$ & $\gamma_{1}=0,\gamma_{2}\neq0$ & $\beta_{3}\neq0$
  & $\gamma_{3}=0$ & $\gamma_{2}\neq0$ & b \\
     & & & $\gamma_{1,2}\neq0$ & & & $\gamma_{1,2}\neq0$ & c \\
  \hline
     & & & $\gamma_{1}\neq0,\gamma_{3}=0$ & & & $\gamma_{1}\neq0$ & a \\
  V & $\alpha_{1,3}\neq0,\alpha_{2}=0$ & $\beta_{1,3}=0$ & $\gamma_{1}=0,\gamma_{3}\neq0$ & $\beta_{2}\neq0$ & $\gamma_{2}=0$
  & $\gamma_{3}\neq0$  & b \\
     & & & $\gamma_{1,3}\neq0$ &  & & $\gamma_{1,3}\neq0$ & c \\
  \hline
     & & & $\gamma_{2}\neq0,\gamma_{3}=0$ & & & $\gamma_{2}\neq0$ & a \\
  VI & $\alpha_{2,3}\neq0,\alpha_{1}=0$ & $\beta_{2,3}=0$ & $\gamma_{2}=0,\gamma_{3}\neq0$ & $\beta_{1}\neq0$
  & $\gamma_{1}=0$ & $\gamma_{3}\neq0$ & b \\
     & & & $\gamma_{2,3}\neq0$ &  & & $\gamma_{2,3}\neq0$& c \\
  \hline
  \hline
\end{tabular}}
\end{table}

\begin{table}[!htbp]\footnotesize
  \centering
  \caption{Classifications of the  constraint conditions on $\kappa_{23}\neq0$ and $\kappa_{12}=\kappa_{13}=0$.}\label{ta:tab2}
  \resizebox{\textwidth}{!}{
  \begin{tabular}{|c|c|c|c|c|c|c|}
  \hline
  \hline
  Entries & $\kappa_{11}\neq0$ & $\kappa_{12}=0$ & $\kappa_{13}=0$ & $\kappa_{22}\neq0$ & $\kappa_{23}\neq0$ & $\kappa_{33}\neq0$ \\
  \hline
   & & & & $\beta_{2}\neq0,\beta_{3}=0$ & $\gamma_{2}\neq0$ & $\gamma_{3}\neq0/=0$ \\
  I& $\alpha_{1}\neq0,\alpha_{2,3}=0$ &  $\beta_{1}=0$& $\gamma_{1}=0$ & $\beta_{2}=0,\beta_{3}\neq0$
  & $\gamma_{3}\neq0$ & $\gamma_{2}\neq0/=0$ \\
   & & & & $\beta_{2,3}\neq0$ & $\gamma_{2}\neq0/\gamma_{3}\neq0/\gamma_{2,3}\neq0$ & $\gamma_{2}\neq0/\gamma_{3}\neq0/\gamma_{2,3}\neq0$ \\
  \hline
     & & & & $\beta_{1}\neq0,\beta_{3}=0$ & $\gamma_{1}\neq0$ & $\gamma_{3}\neq0/=0$ \\
  II & $\alpha_{2}\neq0,\alpha_{1,3}=0$ & $\beta_{2}=0$ &  $\gamma_{2}=0$ & $\beta_{1}=0,\beta_{3}\neq0$
  & $\gamma_{3}\neq0$ & $\gamma_{1}\neq0/=0$ \\
     & & & & $\beta_{1,3}\neq0$ & $\gamma_{1}\neq0/\gamma_{3}\neq0/\gamma_{1,3}\neq0$ & $\gamma_{1}\neq0/\gamma_{3}\neq0/\gamma_{1,3}\neq0$ \\
  \hline
     & & & & $\beta_{1}\neq0,\beta_{2}=0$ & $\gamma_{1}\neq0$ & $\gamma_{2}\neq0/=0$ \\
  III & $\alpha_{3}\neq0,\alpha_{1,2}=0$ & $\beta_{3}=0$ & $\gamma_{3}=0$ & $\beta_{1}=0,\beta_{2}\neq0$
  & $\gamma_{2}\neq0$ & $\gamma_{1}\neq0/=0$ \\
     & & & & $\beta_{1,2}\neq0$ & $\gamma_{1}\neq0/\gamma_{2}\neq0/\gamma_{1,2}\neq0$ & $\gamma_{1}\neq0/\gamma_{2}\neq0/\gamma_{1,2}\neq0$ \\
  \hline
  \hline
  IV & $\alpha_{1,2}\neq0,\alpha_{3}=0$ & $\beta_{1,2}=0$ & $\gamma_{1,2}=0$ & $\beta_{3}\neq0$
  & $\gamma_{3}\neq0$ & $\gamma_{3}\neq0$ \\
  \hline
  V & $\alpha_{1,3}\neq0,\alpha_{2}=0$ & $\beta_{1,3}=0$ & $\gamma_{1,3}=0$ & $\beta_{2}\neq0$ & $\gamma_{2}\neq0$
  & $\gamma_{2}\neq0$  \\
  \hline
  VI & $\alpha_{2,3}\neq0,\alpha_{1}=0$ & $\beta_{2,3}=0$ & $\gamma_{2,3}=0$ & $\beta_{1}\neq0$
  & $\gamma_{1}\neq0$ & $\gamma_{1}\neq0$ \\
  \hline
  \hline
\end{tabular}}
\end{table}

At present we do not have too much knowledge about the absolute neutrino masses and the mass order. Only the two mass differences $\Delta m^{2}_{21}$
and $\Delta m^{2}_{32(31)}$ are well valued. The lightest neutrino mass in normal order (NO) can be vanishing in principle. In the case we may have
the elements $\kappa_{1i}$ ($i=1,2,3$) (and of course $\kappa_{i1}$) are all vanishing ones. Then only the condition $\kappa_{23}\neq0$ is required
to obtain the non-vanishing $\theta_{13}$. The situation is not listed in the tables.

We remind that the scheme is a simplified one for the Majorana mass matrix is just a diagonal one and the relative phases among the
Yukawa entries are also neglected. However, the simple classification can reveal simple relations between Yukawa elements, which may be 
realized in some concrete models. In the following we shall present a discrete flavor symmetry model as one of its implications.

\subsection{Implications}
The neutrino Yukawa matrix $\lambda$ whose elements satisfy the constrained conditions in the tables can be realized in some models based on
discrete family symmetries in principle. According to the conditions that $\alpha_{i}$, $\beta_{i}$, $\gamma_{i}$ are vanishing or not, one may
solve the elements $\lambda_{ij}$ with some private choices. From the above discussions, for the same index $i$, we have divided
$\alpha_{i}$, $\beta_{i}$, $\gamma_{i}$ into eight groups with their values vanishing or not. For example from the case (I.a)
in~\hyperref[ta:tab1]{Table 1} we can obtain the following solutions of nine elements according to the corresponding constrains
\begin{eqnarray}\label{eq:solve}
  & & \alpha_{1}\neq0,\quad\beta_{1}=0,\quad\gamma_{1}\neq0,\Longrightarrow
  \lambda_{11}=\lambda_{1},\quad \lambda_{12}=c_{12}\lambda_{1}, \quad \lambda_{13}=-(1+c_{12})\lambda_{1}, \nonumber\\
  & & \alpha_{2}=0,\quad\beta_{2}\neq0,\quad\gamma_{2}=0,\Longrightarrow
  \lambda_{21}=\lambda_{22}=\lambda_{23}=\lambda_{2}, \nonumber\\
  & & \alpha_{3}=0,\quad\beta_{3}=0,\quad\gamma_{3}\neq0,\Longrightarrow
  \lambda_{31}=0,\quad \lambda_{32}=-\lambda_{33}=\lambda_{3},
\end{eqnarray}
with $c_{12}\neq-\frac{1}{2}$ and $\lambda_{i}\neq0$ ($i=1,2,3$). Then the Yukawa matrix is simply of the texture
\begin{equation}\label{eq:YukawaIa}
  \lambda=\left(
            \begin{array}{ccc}
              \lambda_{1} & c_{12}\lambda_{1} & -(1+c_{12})\lambda_{1} \\
              \lambda_{2} & \lambda_{2} & \lambda_{2} \\
              0 & \lambda_{3} & -\lambda_{3} \\
            \end{array}
          \right).
\end{equation}
In the similar way one may get other possible Yukawa textures. Note that the relation between different $\lambda_{i}$ is undetermined, thus we can
treat them either independent or dependent from each other. Besides, for example, the coefficient $c_{12}$ in eq.~\eqref{eq:YukawaIa} is undetermined
as well. One may note that the entries in Yukawa matrix $\lambda$, while $c_{12}$ is chosen properly, seem to be kind of vev alignment in
some discrete flavor models. Numerous works on the construction of flavor models are proposed by applying discrete non-Abelian family symmetries, one
can see~\cite{Alta/2010,Ish/2010a,King/rev} for review. Taking $c_{12}=-1$ for example (gives $\lambda_{13}=0$), we find that the texture
of $\lambda$ in eq.~\eqref{eq:YukawaIa} can be realized in the $A_{4}$ flavor symmetry model~\footnote{For a concise elucidation,
we suppress the coupling coefficients in front of each Yukawa
coupling term and we also ignore the suppression by powers of
the high energy scale $\Lambda$ in the theory. Besides we also assume
a diagonal mass matrix of charged leptons in such a context.}
\begin{equation}\label{eq:A4model}
  w_{\nu}\supset\nu^{c}_{1}(\ell\phi_{1})''\zeta h_{u}+\nu^{c}_{2}\ell\phi_{2}h_{u}
  +\nu^{c}_{3}\ell\phi_{1}h_{u}+\sum_{i}\frac{1}{2}\mu_{i}\nu^{c}_{i}\nu^{c}_{i},
\end{equation}
where all the fields are assigned to be proper representations under $SU(2)_{L}\times U(1)_{Y}\times A_{4}$
\begin{equation}\label{eq:A4repre}
  \nu^{c}_{1,2,3}\sim(1,0,1),\quad\ell\sim(2,-1,3),\quad\phi_{1,2}\sim(1,1,3),\quad\zeta\sim(1,1,1'),\quad h_{u}\sim(2,1,1).
\end{equation}
The Dirac neutrino mass $m_{D}=\lambda v_{u}$ is achieved by Higgs $h_{u}$ and flavons $\phi_{i}$ obtain their respective vevs. To be specific
$\langle h_{u}\rangle=v_{u}$ is Higgs vev after electroweak symmetry breaking, and after $A_{4}$ symmetry breaking the scalar flavon
fields $\phi_{1,2}$ and $\zeta$ develop their following nontrivial vevs
\begin{equation}\label{eq:flavonvev}
  \langle\phi_{1}\rangle=(0,-1,1)v_{1},\quad\langle\phi_{2}\rangle=(1,1,1)v_{2},\quad \langle\zeta\rangle=v_{\zeta}.
\end{equation}
These vacuum structure can be realized with the flavon fields are charged under some auxiliary Abelian symmetries like $Z_{N}$ or $U(1)_{FN}$,
such as the model~\cite{Ding/2012A4}.

Without entangling with details for the way of produce the vacuum configurations, one may build other
possible Yukawa textures in a non-Abelian discrete family symmetry models. In the case that coefficients $c_{12}$ etc are taken different values,
one vev $\langle\phi_{i}\rangle$ may not be sufficient to produce the expected Yukawa entries, then the desired entries in Yukawa texture can be
obtained by two or more flavons with different vevs combined together. The vevs of flavons may also own other exotic alignment direction.
This would be realized in a specific model by different vacuum alignment mechanism, such as the form dominance~\cite{ChenMC/2009fd} or
constrained sequential dominance~\cite{Antusch/2012csd}, or even taking into account the fine tuning to the coupling parameters is allowed as well.
Moreover it is worth to investigate whether leptogenesis~\cite{Fukugita/1986bg}
works in the background, which is an appealing mechanism to decipher the mystique of the matter-antimatter asymmetry in the universe.
The issues deserve further investigations, which is now beyond the present work and left for future studies.

\section{Conclusion}\label{S4}
The three well determined PMNS leptonic mixing angles especially the last measured mixing angle $\theta^{\ell}_{13}$ ruled out the previous widely
studied mixing patterns $V_{\nu}$ with a vanishing $\theta^{\nu}_{13}$. In this paper we have studied two minimal scenarios that lead to a
nonvanishing $\theta^{\ell}_{13}$. The two scenarios based on the preliminary condition that the effective coupling matrix
$\kappa^{eff}=\lambda^{T}\mu^{-1}\lambda$ is pre-diagonalized by the tribimaximal mixing pattern $V_{\nu}$ with a vanishing $\theta^{\nu}_{13}$ and
large $\theta^{\nu}_{12}$ and $\theta^{\nu}_{23}$. The resulting pre-diagonal symmetrical coupling matrix $\kappa=V^{T}_{\nu}\kappa^{eff}V_{\nu}$
has four texture zeros and only one nonvanishing off-diagonal elements $\kappa_{13}$ or $\kappa_{23}$ in respective scenario. For the case that
all diagonal elements are nonzero we thoroughly classified the constraint conditions which are deduced from
the linear combinations, $\alpha_{i}$, $\beta_{i}$ and $\gamma_{i}$, of Yukawa elements $\lambda_{ij}$ in both scenarios.

By solving a set of the constraint on $\alpha_{1,2,3}$, $\beta_{1,2,3}$ as well as $\gamma_{1,2,3}$, and taking a specific choice of an 
undetermined parameter, we find that the resulting Yukawa matrix $\lambda$ possesses a rather simple texture and can be realized in 
an $A_{4}$ falvor model with some specific vacuum alignment of flavons. The scheme can be also generalized to other contexts, such as minimal 
seesaw with two heavy majorana masses or the case that Majorana mass matrix being diagonalized by the other mixing patterns $V_{\nu}$, etc. 
The vacuum alignments which feature the Yukawa elements can be implemented by specific mechanism or corrections to the original one in 
which fine tuning may be required in coupling coefficients. The resulting Yukawa coupling matrix together with the heavy Majorana masses 
deserve to investigate whether the leptogenesis works in the background.



\begin{thebibliography}{90}
\bibitem{HPS/2002}
P.F.~Harrison, D.H.~Perkins, W.G.~Scott, Phys.~Lett.~\textbf{B 530} (2002) 167; \\
P.F.~Harrison, W.G.~Scott, Phys.~Lett.~\textbf{B 535} (2002) 163.

\bibitem{GR/2007}
Y.~Kajiyama, M.~Raidal, A.~Strumia, Phys.~Rev.~\textbf{D 76} (2007) 117301.

\bibitem{Rode/2009a}
W.~Rodejohann, Phys.~Lett.~\textbf{B 671} (2009) 267.

\bibitem{Adul/2009d}
A.~Adulpravitchai, A.~Blum, W.~Rodejohann, New J.~Phys.~\textbf{11} (2009) 063026.

\bibitem{dyb/2012}
F.P.~An, et al., Daya Bay Collaboration, Phys.~Rev.~Lett.~\textbf{108} (2012) 171803; \\
F.P.~An, et al., Daya Bay Collaboration, Chin.~Phys.~\textbf{C 37} (2013) 011001.

\bibitem{RENO/2012}
J.K.~Ahn, et al., RENO Collaboration, Phys.~Rev.~Lett.~\textbf{108} (2012) 191802.

\bibitem{Frampton/2002tz}
P.H.~Frampton, S. L.~Glashow, and D.~Marfatia, Phys.~Lett.~\textbf{B 536} (2002)~79.

\bibitem{Xing/2002tz}
Z.-z.~Xing, Phys.~Lett.~\textbf{B 530} (2002)~159.

\bibitem{Randhawa/2006tz}
M.~Randhawa, G.~Ahuja, and M.~Gupta, Phys.~Lett.~\textbf{B 643} (2006)~175.

\bibitem{Merle/2006tz}
A.~Merle, and W.~Rodejohann, Phys.~Rev.~\textbf{D 73}(2006)~073012.

\bibitem{DevS/2007tz}
S.~Dev, S.~Kumar, S.~Verma, and S.~Gupta, Phys.~Rev.~\textbf{D 76} (2007)~013002;\\
S.~Dev, S.~Kumar, S.~Verma, and S.~Gupta, Nucl.~Phys.~\textbf{B 784} (2007)~103;\\
G.~Ahuja, S.~Kumar, M.~Randhawa, M.~Gupta, and S.~Dev, Phys.~Rev.~\textbf{D 76} (2007)~013006;\\
S.~Dev, S.~Kumar, Mod.~Phys.~Lett.~\textbf{A 22} (2007)~1401.

\bibitem{Kumar/2011tz}
S.~Kumar, Phys.~Rev.~\textbf{D 84} (2011)~077301.

\bibitem{Ludl/2012tz}
P.O.~Ludl, S.~Morisi, and E.~Peinado, Nucl.~Phys.~\textbf{B 857} (2012)~411.

\bibitem{Grimus/2013tz}
W.~Grimus, and P.O.~Ludl, J.~Phy.~G: Nucl.~Part.~Phys.~\textbf{40}~(2013)~055003.

\bibitem{Blankenburg/2013tz}
G.~Blankenburg, and D.~Meloni, Nucl.~Phys.~\textbf{B 867} (2013)~749.

\bibitem{Fritzsch/2011tz}
H.~Fritzsch, Z.-z.~Xing, and S.~Zhou, JHEP~\textbf{09} (2011)~083.

\bibitem{Hirsch/2007A4}
M.~Hirsch, A.S.~Joshipura, S.~Kaneko, and J.W.F.~Valle, Phys.~Rev.~Lett.~\textbf{99}~(2007)~151802.

\bibitem{ZhouS/2016Tz}
S.~Zhou, Chin.~Phys.~\textbf{C 40} (2016)~033102.

\bibitem{DevS/2012Tz}
S.~Dev, S.~Kumar, S.~Verma, S.~Gupta, and R.R.~Gautam, Eur.~Phys.~J.~\textbf{C 72} (2012)~1940;\\
S.~Dev, L.~Singh, D.~Raj, Eur.~Phys.~J.~\textbf{C 75} (2015)~394.

\bibitem{Deepthi/2012Tz}
K.N.~Deepthi, Srinu Gollu, R.~Mohanta, Eur.~Phys.~J.~\textbf{C 72} (2012)~1888.

\bibitem{LiaoJ/2013Tz}
J.~Liao, D.~Marfatia, K. Whisnant, Nucl.~Phys.~\textbf{B 900} (2015)~449;\\
J.~Liao, D.~Marfatia, K. Whisnant, Phys.~Rev.~\textbf{D 87}~(2013)~073013;\\
J.~Liao, D.~Marfatia, K. Whisnant, Phys.~Rev.~\textbf{D 88}~(2013)~033011.

\bibitem{Ludl/2014sTz}
P.O.~Ludl,  W.~Grimus, Phys.~Lett.~\textbf{B 744}~(2015)~38;\\
P.O.~Ludl and W.~Grimus, JHEP \textbf{07}~(2014)~090.

\bibitem{Berger/2001Tz}
M.S.~Berger, and Kim~ Siyeon, Phys.~Rev.~\textbf{D 64}~(2001)~053006;\\
B.~Adhikary, M.~Chakraborty and A.~Ghosal, Phys.~Rev.~\textbf{D 86}~(2012)~013015;\\
P.M.~Ferreira, and L.~Lavoura,  Mod.~Phys.~Lett.~\textbf{A 27}~(2012)~1250159


\bibitem{Cozero/2013}
S.~Dev, R.R.~Gautam, and L.~Singh, Phys.~Rev.~\textbf{D 87} (2013)~073011.

\bibitem{Lavoura/2005vm}
L.~Lavoura, Phys.~Lett.~\textbf{B 609} (2005) 317.

\bibitem{Lashin/2008vm}
E.I.~Lashin and N.~Chamoun, Phys.~ Rev.~\textbf{D 78} (2008)~073002;\\
E.I.~Lashin and N.~Chamoun, Phys.~ Rev.~\textbf{D 80} (2009)~093004.

\bibitem{DevS/2011vm}
S.~Dev, S.~Gupta, and R.R.~Gautam, Mod.~Phys~Lett.~\textbf{A 26} (2011)~501;\\
S.~Dev, S.~Gupta, R.R.~Gautam, and L.~Singh, Phys.~Lett.~\textbf{B 706} (2011)~168.

\bibitem{ArakiT/2012vm}
T.~Araki, J.~Heeck, and J.~Kubo, JHEP~\textbf{07} (2012)~083.

\bibitem{Verma/2012vm}
S.~Verma, Nucl.~Phys.~\textbf{B 854} (2012)~340.

\bibitem{DevS/2010vm}
S.~Dev, S.~Verma, S.~Gupta, and R.R.~Gautam, Phys.~Rev.~\textbf{D 81} (2010)~053010;\\
S.~Dev, R.R.~Gautam, and L.~Singh, Phys.~Rev.~\textbf{D 89} (2014)~013006.

\bibitem{LiaoJ/2014vm}
J.~Liao, D.~Marfatia, K.~Whisnant, JHEP~\textbf{09} (2014)~013.

\bibitem{KanekoS/2005Hyb}
S.~Kaneko, H.~Sawanaka, and M.~Tanimoto, JHEP~\textbf{08} (2005)~073.

\bibitem{DevS/2010Hyb}
S.~Dev, S.~Verma, and S.~Gupta, Phys.~Lett.~\textbf{B 687} (2010)~53;\\
S.~Dev, S.~Gupta, and R.R.~Gautam, Phys.~Rev.~\textbf{D 82} (2010)~073015.

\bibitem{Goswami/2010Hyb}
S.~Goswami, S.~Khan, and A.~Watanabe, Phys.~Lett.~\textbf{B 693} (2010)~249.

\bibitem{Garcia/2015fit}
M.C.~Gonzalez-Garcia, M.~Maltoni, T.~Schwetz, Nucl.~Phys.~\textbf{B 908} (2016)~199.

\bibitem{Alta/2010}
G.~Altarelli, F.~Feruglio, Rev.~Mod.~Phys.~\textbf{82} (2010) 2701.

\bibitem{Ish/2010a}
H.~Ishimori, T.~Kobayashi, H.~Ohki, Y.~Shimizu, H.~Okada, M.~Tanimoto, Prog.~Theor.~Phys.~Suppl.~\textbf{183} (2010) 1-163.

\bibitem{King/rev}
S.F.~King, C.~Luhn, Rep.~Prog.~Phys.~\textbf{76} (2013) 056201;\\
S.F.~King, A.~Merle, S.~Morisi, Y.~Shimuzu, M.~Tanimoto, New.~J.~Phys.~\textbf{16} (2014) 045018;\\
S.F.~King, J.~Phys.~G: Nucl.~Part.~Phys. \textbf{42} (2015) 123001.

\bibitem{Ding/2012A4}
G.-J.~Ding, D.~Meloni, Nucl.~Phys.~\textbf{B 855} (2012)~21.

\bibitem{ChenMC/2009fd}
M.C.~Chen, S.F.~King, JHEP~\textbf{06} (2009)~072.

\bibitem{Antusch/2012csd}
S.~Antusch, S.F.~King, M.~Spinrath, Nucl.~Phys.~\textbf{B 856} (2012)~328.

\bibitem{Fukugita/1986bg}
M.~Fukugita, T.~Yanagida, Phys.~Lett.~\textbf{B~174}~(1986)~45;\\
S.~Davidson,E.~Nardiand, Y.~Nir, Phys.~Rept.~\textbf{105}~(2008)~466.







\end{thebibliography}
\end{document}